\documentclass[aps,prl,twocolumn,groupedaddress,draft,showpacs,nofootinbib]{revtex4}
\usepackage{graphicx,epsf}
\newif\ifdraft
\drafttrue
\newif\ifpreprint
\preprinttrue

\def\fig#1{fig.~{\ref{#1}}}
\def\Fig#1{Fig.~{\ref{#1}}}

\def\spa#1.#2{\left\langle#1\,#2\right\rangle}
\def\spb#1.#2{\left[#1\,#2\right]}
\def\tree{{\rm tree}}

\def\eps{\epsilon}

\def\nn{\nonumber}

\def\eqn#1{eq.~(\ref{#1})}

\def\NeqOne{{{\cal N}=1}}

\def\NeqFour{{{\cal N}=4}}

\def\NeqEight{{{\cal N}=8}}

\def\be{\begin{equation}}
\def\ee{\end{equation}}
\def\bea{\begin{eqnarray}}
\def\eea{\end{eqnarray}}
\def\ba{\begin{eqnarray}}
\def\ea{\end{eqnarray}}

\def\tree{{\rm tree}}

\def\fourloop{{\rm 4\hbox{-}loop}}

\newbox\charbox
\newbox\slabox
\def\s#1{{      % Feynman slash
        \setbox\charbox=\hbox{$#1$}
        \setbox\slabox=\hbox{$/$}
        \dimen\charbox=\ht\slabox
        \advance\dimen\charbox by -\dp\slabox
        \advance\dimen\charbox by -\ht\charbox
        \advance\dimen\charbox by \dp\charbox
        \divide\dimen\charbox by 2
        \raise-\dimen\charbox\hbox to \wd\charbox{\hss/\hss}
        \llap{$#1$} }}

\begin{document}

%\ifpreprint
\hbox{UCLA/09/TEP/09/47
\hskip 11.4 cm  SLAC--PUB--13608}
%\fi

\title{The Ultraviolet Behavior of $\NeqEight$ Supergravity at Four
      Loops}

\author{Z.~Bern${}^a$, J.~J.~Carrasco${}^a$, L.~J.~Dixon${}^{b}$,
 H.~Johansson${}^a$, 
 and  R.~Roiban${}^c$ }
\affiliation{
${}^a$Department of Physics and Astronomy, UCLA, Los Angeles, CA
90095-1547, USA  \\
${}^b$SLAC National Accelerator Laboratory,
              Stanford University,
             Stanford, CA 94309, USA \\
${}^c$Department of Physics, Pennsylvania State University,
           University Park, PA 16802, USA
}

\begin{abstract}
We describe the construction of the complete four-loop four-particle
amplitude of $\NeqEight$ supergravity.  The amplitude is ultraviolet
finite, not only in four dimensions, but in five
dimensions as well.  The observed extra cancellations provide additional
non-trivial evidence that $\NeqEight$ supergravity in four dimensions
may be ultraviolet finite to all orders of perturbation theory.
\end{abstract}

\pacs{04.65.+e, 11.15.Bt, 11.25.Db, 12.60.Jv \hspace{1cm}}

\maketitle

An often-expressed sentiment is that point-like quantum field theories
based on Einstein's theory of General Relativity, including
supersymmetric extensions thereof, are quantum mechanically
inconsistent, due to either a proliferation of divergences associated
with the dimensionful nature of Newton's constant, or absence of
unitarity.  A series of recent computations has challenged this widely
held belief.  In particular, the three-loop four-graviton
amplitude~\cite{GravityThree,CompactThree} in $\NeqEight$
supergravity~\cite{CremmerJulia} exposes cancellations beyond those
needed for ultraviolet (UV) finiteness at that order.  Novel
cancellations occur already in this theory~\cite{NoTriangle,
VanhoveOrderless} at one loop, related~\cite{UnexpectedOneloop,
Simplest} to the remarkably good behavior of gravity tree amplitudes
under large complex deformations of external
momenta~\cite{GravityLargeZ, Simplest}, and to the unordered nature of
gravity amplitudes~\cite{VanhoveOrderless}.  The modern unitarity
method~\cite{UnitarityMethod} implies that extensive UV cancellations
occur to {\it all} loop orders~\cite{Finite}, for a class of terms
obtained by isolating one-loop sub-amplitudes via generalized
unitarity~\cite{GeneralizedUnitarity}, leading to the 
proposal~\cite{UnexpectedOneloop} that the multiloop UV cancellations
trace back to the tree-level behavior.
These surprising cancellations
point to the possible perturbative UV finiteness of the theory.

Interestingly, M theory and string theory have also been used to argue
both in favor of the finiteness of $\NeqEight$
supergravity~\cite{DualityArguments}, and that divergences are delayed
through nine loops~\cite{Berkovits,GreenII}; issues involving the
decoupling of certain massive states~\cite{GOS} remain in either
case.  The non-compact $E_{7(7)}$ duality symmetry 
of $\NeqEight$ supergravity~\cite{CremmerJulia,E7Recent}
may also play a role~\cite{Simplest,KalloshLightCone},
though this remains to be demonstrated.
A mechanism rendering a point-like theory of quantum gravity
ultraviolet finite would be novel and should have a profound impact on
our understanding of gravity.

Indeed, all studies to date conclude that supersymmetry and gauge
invariance alone cannot prevent the onset of UV divergences to all
loop orders in four dimensions. In fact, it had been a longstanding
expectation that, in generic supergravity theories, four-graviton
amplitudes diverge at three loops in four
dimensions~\cite{Supergravity}.  Such a divergence would be associated
with a counterterm composed of four appropriately contracted Riemann
tensors (the square of the Bel-Robinson tensor), denoted by $R^4$.  A
recent study~\cite{HoweStelleRecent} explains the known lack of this
counterterm~\cite{GravityThree,CompactThree}, both in terms of
non-renormalization theorems and an algebraic formalism for
constraining counterterms.  However, it does predict divergences 
at $L=5$ loops in dimension $D=4$ and at $L=4$ loops in
$D=5$~\footnote{Higher-dimensional maximal supergravities
may be understood as dimensional reductions of $\NeqOne$ supergravity
in $D = 11$.}, unless additional cancellation mechanisms beyond
supersymmetry and gauge invariance are present.

In contrast, explicit computations of the four-graviton amplitude at
successive loop orders have consistently revealed unexpected UV
cancellations.  Results at two loops strongly suggested~\cite{BDDPR},
and at three loops proved~\cite{GravityThree,CompactThree} that the
$R^4$ divergence is absent in $\NeqEight$ supergravity. In addition,
UV divergences are absent at three loops in $D<6$.  The theory
first diverges in $D=6$, and the counterterm has the schematic form 
${\cal D}^6 R^4$, where ${\cal D}$ is a space-time derivative
acting on the Riemann tensors~\cite{CompactThree}.
The computation described in this letter reveals no UV divergences 
at four loops in both $D=4$ and $D=5$, specifically ruling out
a counterterm of the form ${\cal D}^6 R^4$ in $D=5$.
The origin of the observed UV properties is, however, not yet
properly understood.  

It is worth noting that more speculative field-theoretic
studies have suggested
further delays to the onset of divergences.  For example, if off-shell
superspaces with manifest ${\cal N} = 6,7$ or $8$ supersymmetries were
to exist, $D=4$ divergences would be delayed to at least
$L=5,6$ or $7$ loops, respectively~\cite{GrisaruSiegel,HoweStelleNew}.
Locality of counterterms in ${\cal N}=8$
light-cone superspace has also been used to
argue~\cite{KalloshLightCone} for an $L=7$ bound.
With the additional speculation that all fields
respect an eleven-dimensional gauge symmetry, one can even delay the first
potential divergence to nine loops~\cite{HoweStelleRecent}. Interestingly,
this bound coincides with the one suggested~\cite{GreenII} from a
string theory non-renormalization theorem~\cite{Berkovits}.

%%%%%%%%% FIGURE %%%%%%%%%%%%%%%
\begin{figure}[t]
\centerline{\epsfxsize 3.3 truein \epsfbox{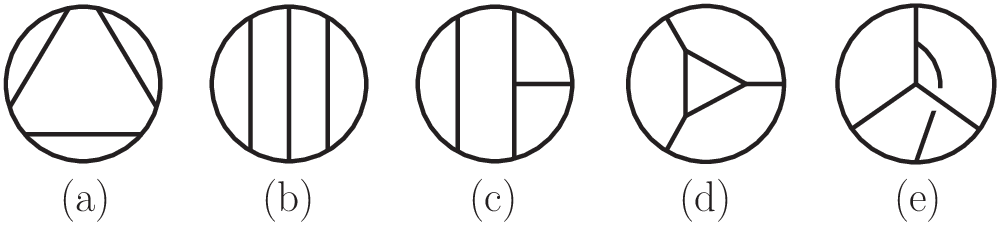}}
\caption[a]{\small Vacuum graphs from which one can build the
contributing four-point graphs by attaching external legs.  They are
also useful for classifying the UV divergences.}
\label{VacuumTopologyFigure}
\end{figure}
%%%%%%%%%%%%%%%%%%%%%%%%%%%%%%%%

%%%%%%%%% FIGURE %%%%%%%%%%%%%%%
\begin{figure}[t]
\centerline{\epsfxsize 3.4 truein \epsfbox{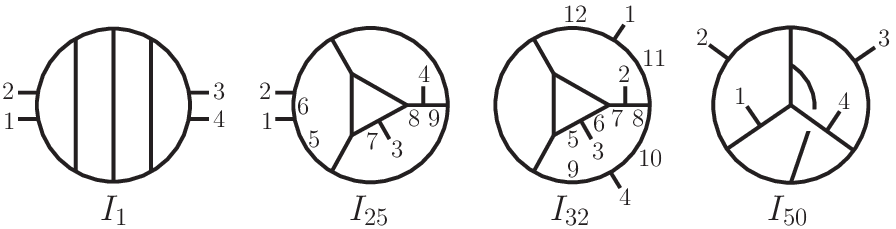}}
\caption[a]{\small Four of the 50 distinct graphs corresponding to the
integrals composing the result for $M_4^\fourloop$.  }
\label{ExampleL8Figure}
\end{figure}
%%%%%%%%%%%%%%%%%%%%%%%%%%%%%%%%

In this letter, we describe the four-loop four-particle amplitude of 
$\NeqEight$ supergravity, denoted by $M_4^\fourloop$, 
which we represent as a sum of 50 four-loop integrals $I_i$,
\begin{eqnarray}
M_4^{\fourloop} \! & = & \! -\Bigl({\kappa \over 2}\Bigr)^{10}
  \! s t u M_4^\tree  \sum_{S_4}\,
\sum_{i=1}^{50} c_i  I_i \,.  \hskip .3 cm 
\label{FourLoopAmplitude}
\end{eqnarray}
Here $S_4$ is the set of 24 permutations of the massless external legs
$\{1,2,3,4\}$ with momenta $k_i$, the $c_i$ are combinatorial 
factors depending on the symmetries of the integrals, 
$\kappa$ is the gravitational coupling, 
and $M_4^\tree$ is the corresponding four-point tree
amplitude.  (All 256$^4$ four-point amplitudes of $\NeqEight$ 
supergravity are related to each other by supersymmetry, which
enforces the proportionality of $M_4^{\fourloop}$ to its tree-level
counterpart $M_4^\tree$.)  
The Mandelstam invariants are $s = (k_1 + k_2)^2$,
$t=(k_2+k_3)^2$, $u=(k_1+k_3)^2$.  Each integral $I_i$ corresponds to
a four-loop graph with 13 propagators and 10 cubic vertices. 
The 50 graphs may be obtained by attaching four
external legs to the edges of the five vacuum graphs in
\fig{VacuumTopologyFigure}. Not all possibilities contribute, however;
diagrams containing nontrivial two- or three-point
subgraphs, such as all those obtained from
\fig{VacuumTopologyFigure}(a), do not appear in the amplitude.
Every integral takes the form
\begin{equation}
I_i = \int  \Biggl[ \prod_{p=1}^4 \frac{d^D l_{n_p}}{(2\pi)^D} \Biggr] \, 
 \frac{N_i (l_j, k_j)} {\prod_{n=1}^{13} l_n^2} \,,
\label{IntegralNormalization}
\end{equation}
where the propagator momenta $l_n$ are linear combinations of
four independent loop momenta $l_{n_p}$ and the
external momenta $k_j$.
The numerator polynomial
$N_i(l_j, k_j)$ is of degree 12 in the momenta, by dimensional
analysis.  Generically, we denote loop momenta by $l$ and external momenta
by $k$.

The full amplitude is too lengthy to present in this letter.  Rather,
we outline its construction and demonstrate some of the relevant UV
cancellations.  Explicit expressions for the numerators, symmetry
factors and propagators may be found online~\cite{WebResult}.  As
examples, the graphs for four integrals, labeled $I_1$, $I_{25}$,
$I_{32}$ and $I_{50}$ in the online expressions~\cite{WebResult}, are
shown in~\fig{ExampleL8Figure}.

\begin{figure}[t]
\centerline{\epsfxsize 3.35 truein \epsfbox{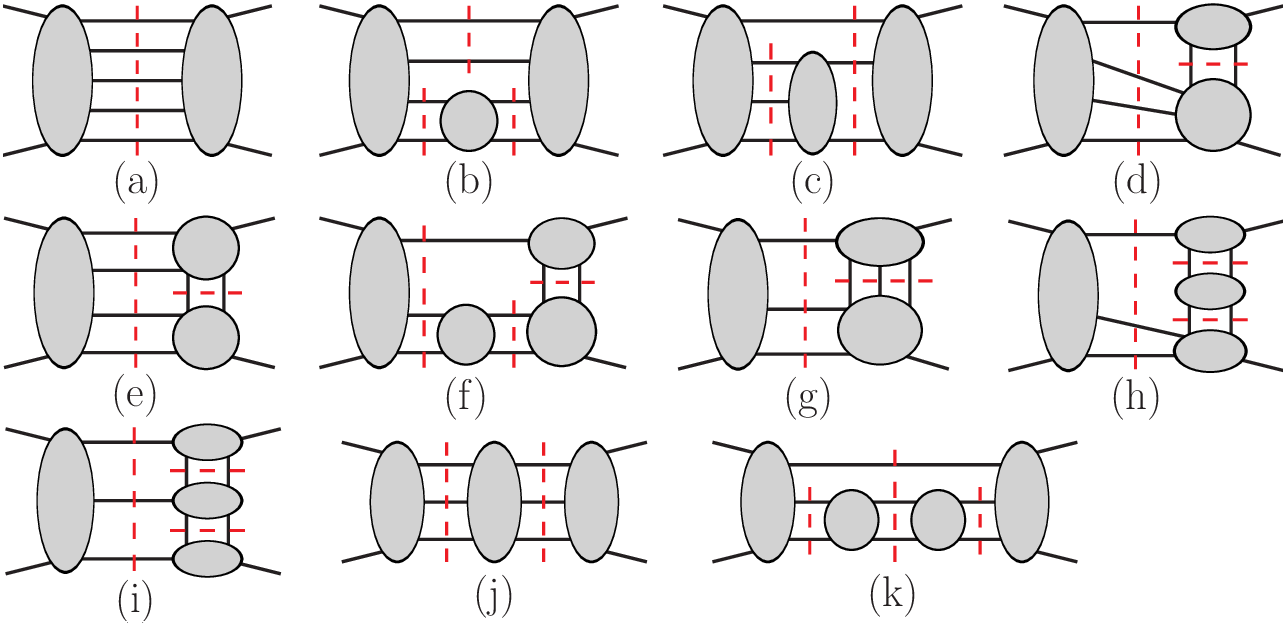}}
\caption[a]{\small Evaluating these 11 cuts, along with 15
two-particle reducible cuts, suffices to uniquely
determine the four-loop four-point amplitude. Each blob
denotes a tree amplitude. }
\label{CutBasisFigure}
\end{figure}

To determine the amplitude, we first construct an ansatz with
numerator polynomials $\widetilde N_i(l_j, k_j)$ that contain undetermined
coefficients.  Then we consider generalized unitarity cuts
decomposing the four-loop amplitude into a product of tree
amplitudes $M^\tree_{(i)}$, as shown in \fig{CutBasisFigure}.
Equating the cuts of the ansatz to
the corresponding cuts of the amplitude,
\begin{equation}
M_4^{\fourloop} \Bigr|_{\rm cut}
= \sum_{\rm states} M^\tree_{(1)} M^\tree_{(2)} \cdots  M^\tree_{(n)} \,,
\label{GenCut}
\end{equation}
constrains the undetermined coefficients in the ansatz.

As only tree amplitudes enter \eqn{GenCut}, we follow the 
strategy~\cite{BDDPR} of re-expressing the $\NeqEight$ supergravity cuts in
terms of sums of products of related cuts of the four-loop four-gluon
amplitude in $\NeqFour$ super-Yang-Mills (sYM)
theory~\cite{FourLoopPlanarYM, FourLoopYM}.  The strategy relies on
the Kawai-Lewellen-Tye (KLT) relations between gravity and gauge
theory tree amplitudes~\cite{KLT}, facilitated by their recent
reorganization in terms of diagrams~\cite{TreeJacobi}.  While we
suspect that a representation of the $\NeqEight$ amplitude exists in
which each $N_i$ is at most of degree four in the loop momenta, it is
natural, given the squaring nature of the KLT relations, to first
solve the cut constraints with this condition relaxed.  We
present a solution in which each $N_i$ is at most of degree
eight~\cite{WebResult}.  This representation is sufficient for our
purpose of demonstrating UV finiteness in $D=4,5$.

The KLT relations are valid in arbitrary
dimensions.  Thus, if the $\NeqFour$ amplitudes are valid in $D$
dimensions, then so are the $\NeqEight$ amplitudes derived from them.
While we do not yet have a complete proof of the $(D>4)$-dimensional
validity of the non-planar contributions to the four-loop $\NeqFour$
amplitudes, we have carried out extensive checks.  In particular, 
ordinary two-particle cuts and cuts isolating four-point subamplitudes
extend easily to $D$ dimensions~\cite{FourLoopPlanarYM,
FourLoopYM, TreeJacobi}.  The full $\NeqFour$ sYM
amplitude, the details of its calculation, and non-trivial
consistency checks will be presented elsewhere~\cite{FourLoopYM}.

%%%%%%%%% FIGURE %%%%%%%%%%%%%%%
%%%%%%%%%%%%%%%%%%%%%%%%%%%%%%%%

Following the method of maximal cuts~\cite{FiveLoop,CompactThree}, we
first fix those coefficients of the $\widetilde N_i(l_j,k_j)$ 
that contribute when the number of cut propagators is
maximal---13 in this case. We then
consider cuts with 12 cut lines, fixing the coefficients that appear
in terms proportional to single inverse propagators $l_n^2$ ({\it i.e.},
contact terms).  We continue this procedure down to nine cut lines,
considering, in total, $2\,906$ distinct cuts.  At this
point, the resulting expression is complete, which we demonstrate using a
set of 26 cuts, sufficient to completely determine any four-loop
four-point amplitude in any massless theory.  
The 11 cuts that cannot be straightforwardly verified using
lower-loop four-point amplitudes in two-particle cuts are shown in 
\fig{CutBasisFigure}.

The UV properties of the amplitude are determined by the
numerator polynomials $N_i$.  We decompose them into 
expressions $N_i^{(m)}$ containing all terms with $m$ powers of
loop momenta (and ${12-m}$ powers in the external momenta),
\begin{equation}
N_i = N_i^{(8)} + N_i^{(7)}  + N_i^{(6)}+ \ldots + N_i^{(0)} \,.
\end{equation}
There is some freedom in this decomposition, including
that induced by the choice of independent $l_{n_p}$ 
in the loop integral~(\ref{IntegralNormalization}). 
The overall scaling behavior of
\eqn{IntegralNormalization} implies that an integral with $N_i^{(m)}$
in the numerator is finite when $4 D - 26 + m <0$.  For $m$ odd, by
Lorentz invariance, the leading divergence trivially vanishes under
integration, effectively reducing $m$ by one.  Our representation has
$m\le 8$ for all terms; hence the four-loop amplitude is manifestly UV
finite in $D=4$.

Demonstrating UV finiteness in $D=5$ is more subtle.  It
requires the cancellation of divergences for $m=6,7,8$.  
We employ a systematic procedure for extracting divergences 
from multiloop integrals by expanding in small external
momenta~\cite{MarcusSagnotti}.

We find that the numerator terms with $m=8$ can all
be expressed solely in terms of inverse propagators $l_n^2$;
those with $m=7$ have six powers of loop momenta carried by 
inverse propagators; and those with $m=6$ have four powers; 
schematically,
\begin{eqnarray}
&& \hskip -.7cm 
N_i^{(8)} \sim s_a s_b  l_j^2 l_n^2 l_p^2 l_q^2 \,, \nn \\
&& \hskip -.7 cm 
N_i^{(7)} \sim s_a s_b (k_j\cdot l_n) l_p^2 l_q^2 l_r^2 \,,  \\
&& \hskip -.7 cm 
N_i^{(6)} \sim s_a s_b (k_j\cdot l_n) (k_p\cdot l_q) l_r^2 l_w^2 +
s_a s_b s_c  (l_j\cdot l_n)  l_p^2 l_q^2 \,, \nn
\end{eqnarray}
where each $s_a$ denotes $s$, $t$ or $u$.
After expanding
in small external momenta, potential UV divergences enter through
vacuum integrals, just as at three loops~\cite{GravityThree}.
Vacuum integrals also exhibit infrared
singularities, which we regularize
by injecting two fictitious off-shell external momenta at appropriate
locations in the graph.

Only 12 of the 50 integrals have a nonvanishing $N_i^{(8)}$; all of
them are associated with vacuum diagrams (d) and (e) of
\fig{VacuumTopologyFigure}.
For example, the $k^4 l^8$ terms in the numerators of  
the integrals $I_{25}$ and $I_{32}$ in \fig{ExampleL8Figure} are
\begin{eqnarray}
% diagram 25
N_{25}^{(8)}
&=& \frac{1}{8} l^2_{5} l^2_{6} l^2_{7} 
 \Bigl[ (30 s^2 + 13 t^2 + 13 u^2)l^2_{9}  \nn \\ 
&& \hskip 1.1 cm \null
-(32 s^2 + 19 t^2 + 19 u^2) l^2_{8} \Bigr] \,, 
\nn \\ 
% diagram 32
N_{32}^{(8)} 
&=& \frac{1}{8} \Bigl\{ 2 (7 s^2 + 7 t^2 + 6 u^2) 
               l^2_{5} l^2_{8} l^2_{10} l^2_{12} \nn \\ 
&& \hskip.1cm \null
+ l^2_{9} \Bigl[ 
12 (2 s^2 - t^2 + 2 u^2) l^2_{6} l^2_{7} l^2_{12} \nn \\ 
&& \hskip.8cm \null
- (24 s^2 + 19 t^2 + 19 u^2) l^2_{5} l^2_{8} l^2_{11} \Bigr] \Bigr\}
\,.  \hskip .5 cm 
\label{N_25_32_8}
\end{eqnarray}
All of the $l_n^2$ factors in \eqn{N_25_32_8} cancel
propagators in the integrals. Thus, to leading order in the expansion 
in small external momenta, the $k^4 l^8$ terms in $I_{25}$ and $I_{32}$
reduce to the vacuum diagram $V^{\rm (d)}$ of
\fig{VacuumTopologyFigure}(d),
\begin{eqnarray}
&&
I_{25}\rightarrow -14 (s^2+t^2+u^2) V^{\rm (d)} +O(k^5) \,, \nn \\ &&
I_{32}\rightarrow +14 (s^2+t^2+u^2) V^{\rm (d)}+O(k^5) \,.
\end{eqnarray}
Here we have summed over the $S_4$ permutations of external legs in
\eqn{FourLoopAmplitude}. Because their combinatorial factors $c_{25}$ 
and $c_{32}$ are equal~\cite{WebResult},
the $I_{25}$ and $I_{32}$ contributions cancel at leading order.
Similarly, all $k^4 l^8$
contributions in the remaining diagrams cancel, independent of $D$.

As the $k^5 l^7$ terms cannot generate a leading divergence, 
we need only inspect the $k^6 l^6$ term to determine the 
UV properties of the amplitude in $D=5$.
It is necessary
to expand all integrands down to $k^6 l^6$.  For the 12
integrals starting at $O(k^4 l^8)$, two derivatives are required
with respect to the external momenta $k_i$, acting on propagators of
the form $1/(l_j+K_n)^2$ (where $K_n$ denotes a sum of external
momenta).  The numerators obtained by expanding the integrals to this
order have the schematic form,
\begin{equation}
N_i^{(6)} + N_i^{(7)} { K_n\cdot l_j \over l_j^2}
+ N_i^{(8)}\biggl( 
{ K_n^2 \over l_j^2}
+ { K_n\cdot l_j \,\, K_q\cdot l_p \over l_j^2 l_p^2 } \biggr) \,.
\label{L6expansion}
\end{equation}
The additional denominators can lead to doubled or even tripled
propagators for the graphs in \fig{VacuumTopologyFigure}.
Vacuum integrals with $l^\mu_i l^\nu_j$ in the numerator
can be reduced using Lorentz invariance, 
$l^\mu_i l^\nu_j \rightarrow \eta^{\mu \nu} l_i \cdot l_j/D $, with
$D=5$.  After this reduction, the potential UV divergence is
described by 30 vacuum integrals.
Of these, 23 possess no loop momenta in the numerator, while seven
have an $(l_i+l_j)^2$ numerator factor that cannot be reduced to
inverse propagators using momentum conservation.  There are many ways
to expand the original 50 integrals $I_i$.  Shifting the loop momenta
in \eqn{IntegralNormalization} by
$d^Dl_{n_p} \rightarrow d^D(l_{n_p}+k_j)$
leads to different representations of the terms proportional to
$N_i^{(7)}$ and $N_i^{(8)}$ in \eqn{L6expansion}, and hence to
different forms of the UV divergences in terms of the 30 vacuum
integrals.  Requiring that the different forms are equal generates
identities between vacuum integral divergences.  These identities
suffice to demonstrate cancellation of the $k^6l^6$ divergence in
$M_4^\fourloop$.

Independently, we verified the identities by evaluating all
30 vacuum integrals analytically in $D=5-2\eps$.  To do this we
injected external off-shell momenta and factorized the resulting four-loop
propagator integrals into the product of one-loop and three-loop
propagator integrals, much as we did at three
loops~\cite{CompactThree}.  Integration by parts~\cite{IBP} was used
to reduce the three-loop propagator integrals to master integrals.

\begin{figure}[t]
\centerline{\epsfxsize 3.4 truein \epsfbox{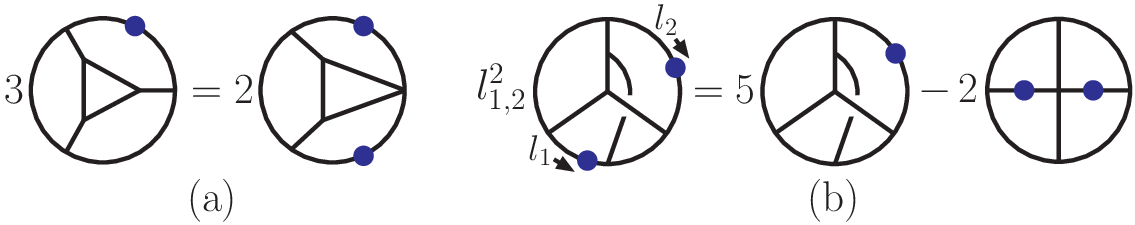}}
\caption[a]{\small Two of the vacuum relations used to analyze
the $D=5$ divergence.  They are valid in $D=5-2\eps$ to order $1/\eps$.
Dots denote doubled propagators and $l^2_{1,2}=(l_1+l_2)^2$ 
represents a numerator factor inside the integral.}
\label{VacuumRelationsFigure}
\end{figure}

Both the vacuum integral identities and the direct integral evaluation
lead to the exact cancellation of the potential $D = 5$ UV divergence.
It is striking that this cancellation can be demonstrated using only
the consistency of the small momentum expansion.
\Fig{VacuumRelationsFigure} displays two of the 16 vacuum integral
identities needed to demonstrate 
finiteness in $D=5$.  

%%%%%%%%% FIGURE %%%%%%%%%%%%%%%
%%%%%%%%%%%%%%%%%%%%%%%%%%%%%%%%

The $k^6 l^6$ cancellation rules out a ${\cal D}^6R^4$ counterterm
in $D=5$.  It implies that the first potential divergence
is proportional to $k^8$ (since a divergence must have an
even power of $k$), corresponding to $D=11/2$.  As the four-loop
four-point $\NeqFour$ sYM amplitude diverges in
$D=11/2$~\cite{FourLoopPlanarYM,FourLoopYM}, the corresponding
$\NeqEight$ supergravity amplitude behaves no worse.

In summary, the results presented here demonstrate that the four-loop
four-particle amplitude of $\NeqEight$ supergravity is UV finite
in $D < 11/2$.  Finiteness in $5\le D<11/2$ is a consequence of
nontrivial cancellations, beyond those already found at three
loops~\cite{GravityThree,CompactThree}. 
From a traditional vantage point of
supersymmetry~\cite{Supergravity,HoweStelleNew,HoweStelleRecent}, our
results are surprising and lend additional support to the possibility
that $\NeqEight$ supergravity is a perturbatively consistent quantum
theory of gravity. 

We thank Sergio Ferrara, Paul Howe, Harald Ita, David Kosower and
Kelly Stelle for helpful discussions and encouragement.  This research
was supported by the US DOE under contracts DE--FG03--91ER40662,
DE-AC02-76SF00515, DE--FG02--90ER40577 (OJI) and the US NSF under
grant PHY-0608114.  R.~R.  acknowledges support by the A.~P. Sloan
Foundation.  J.~J.~M.~C. and H.~J. gratefully acknowledge the
financial support of Guy Weyl Physics and Astronomy Alumni Grants.

%%%%%%%%%%%%%%%%%%%%%%%%%%%%%%%%%%%%%%%%%%%%%%%%%%%%%%%%%%%%%%%%%%%%%%

\end{document}